\documentclass[10pt,journal]{IEEEtranTCOM}

\usepackage[english]{babel}
\usepackage[usenames]{color}
\usepackage[cp1250]{inputenc}
\usepackage{amsfonts}
\usepackage{amsthm}
\usepackage{graphicx}
\usepackage{epsfig}
\usepackage{mathrsfs}
\usepackage{amsmath}
\usepackage{algorithm}
\usepackage{algorithmic}
\usepackage{hyperref}

\theoremstyle{plain}

\begin{document}
\newcommand{\bea}{\begin{eqnarray}}
\newcommand{\eea}{\end{eqnarray}}
\newcommand{\be}{\begin{equation}}
\newcommand{\ee}{\end{equation}}
\newcommand{\beas}{\begin{eqnarray*}}
\newcommand{\eeas}{\end{eqnarray*}}
\newcommand{\bs}{\backslash}
\newcommand{\bc}{\begin{center}}
\newcommand{\ec}{\end{center}}
\def\SC {\mathscr{C}}

\title{Two-way quantum computers\\ adding CPT analog of state preparation}
\author{\IEEEauthorblockN{Jarek Duda}\\
\IEEEauthorblockA{Jagiellonian University,
Lojasiewicza 6, 30-348 Krakow, Poland,
Email: \emph{dudajar@gmail.com}}}
\maketitle

\begin{abstract}
Standard one-way quantum computers (1WQC) combine time symmetric unitary evolution, with asymmetric treatment of boundaries: state preparation allows to enforce a chosen initial state, however, for the final state measurement chooses a random value instead. As e.g. pull/push, negative/positive pressure, stimulated emission/absorption causing deexcitation/excitation are CPT analogs, and one can be used for state preparation, the second should allow for its CPT analog, referred here as CPT(state preparation) - allowing for additional chosen enforcement of the final state, its more active treatment than measurement. It should act similarly to postselection, but through applied physical constraints (instead of running multiple times). Like pumped to $|1\rangle$ prepared state vs its "unpumped" $\langle 0|$ CPT analog, hopefully allowing to construct two-way quantum computers (2WQC) e.g. hydrodynamical, and hopefully photonic: seen as $\langle \Phi_{\textrm{final}}|U_{\textrm{quantum gates}}|\Phi_{\textrm{initial}}\rangle$ like for scattering matrix, with influenced both initial and final states. If possible, for example for an instance of 3-SAT problem on $n$ variables, we could prepare ensemble of $2^n$ inputs with Hadamard gates, calculate 3-SAT alternatives for them, and use CPT(state preparation) to enforce outcomes of all these alternatives to '1'. This way hopefully restricting this ensemble to satisfying given 3-SAT problem: $\sum_{a:\textrm{SAT}(a)} |a\rangle$, in theory allowing to attack NP problems by simultaneously pushing and pulling information through the system for better control.
\end{abstract}
\section{Introduction}
\begin{figure}[t!]
    \centering
        \includegraphics[width=9cm]{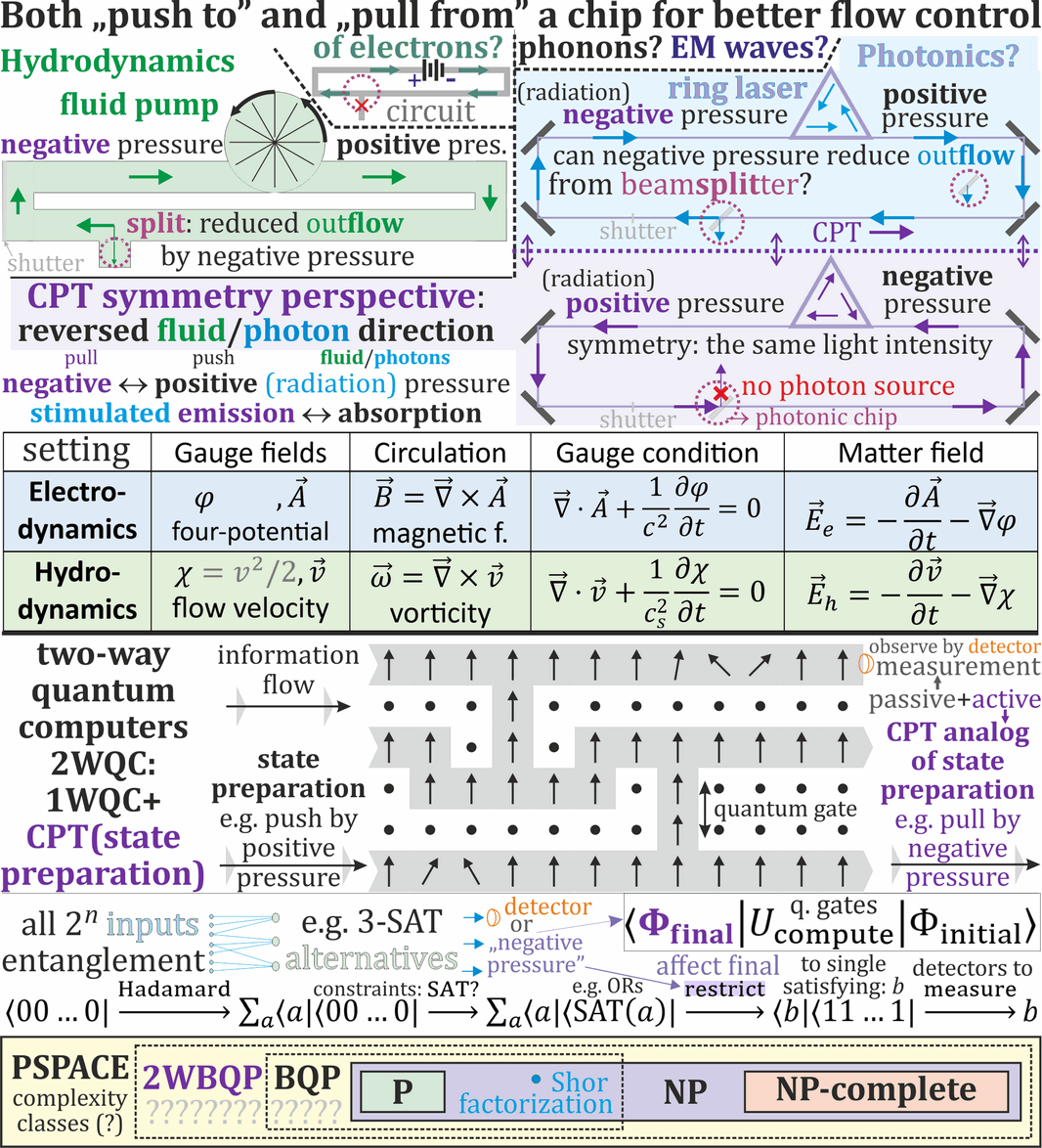}
        \caption{While standard one-way quantum computers (1WQC)~\cite{1WQC} enforce only the initial state intuitively "pushing" information into a (quantum) system, connecting it to some kind of pump would allow to simultaneously "pull" to enforce also the final state in 2WQC - control the information flow actively from both sides.  
        \textbf{Top}: basic scenario with some kind of negative pressure reducing flow down the marked splits - natural for fluid (left), also electrons in conductor cable in external electric field (top). Using mathematical similarity between hydrodynamics and electromagnetism, e.g. in the shown table from \cite{EMh}, we will argue possibility of EM realization of analogous situations especially in photonics quantum computer~(\cite{qc1,qc2,qc3,qc4}) settings. Using stimulated emission-absorption as CPT analogs suggests ring laser should act as such pump for photons. Looking at such setting from perspective after CPT symmetry, photon trajectories would be reversed, for symmetric laser photon number should be the same but in the opposite direction - suggesting the beamsplitter should be practically ignored thanks to simultaneous positive-negative pressure. 
        Alternatively, there are also quantum computer approaches using e.g. superfluids~\cite{superfluid}, mechanical phonons~\cite{phonon}, many use microwaves~\cite{micro} - different technologies can be considered.
        \textbf{Bottom}: being able to actively affect quantum chip in both directions as $\langle \Phi_{\textrm{final}}|U_{\textrm{quantum gates}}|\Phi_{\textrm{initial}}\rangle$ like above (e.g. in superfluid, maybe phonon, microwave, photonic, or other approaches), in theory could allow to attack NP complete problems like 3-SAT. For example similarly to Shor algorithm: prepare ensemble of all inputs, calculate classical function for them - here e.g. alternatives of 3-SAT problems. Then we would like enforce '1' for all these alternatives by such CPT(state preparation) - hopefully restricting ensemble to $\sum_{a:\textrm{SAT}(a)} |a\rangle$, measuring of which should provide solution. Standard 1WQC achieve BQP complexity class. Defining analogously 2WBQP for 2WQC, it would contain NP problems, however, its exact situation between NP and PSPACE is a difficult open problem. }       \label{intr}
\end{figure}

Hydrodynamics and electrodynamics are governed by similar wavelike mathematics, as in Table in Fig. \ref{intr} from \cite{EMh}. For the former there are used microfluidic chips~\cite{fluid} with pump to push fluid inside with positive pressure, analogously in the latter there are e.g. (quantum) photonic chips pushing photons inside with laser. Imagining flow outside can be through one of two pipes, we could \textbf{passively observe/measure} from which one, but can also \textbf{actively pull with negative pressure} for better flow control. This article discusses such active treatment as additional operation for quantum computers (e.g. photonic), using CPT analog of process used for state preparation, potentially allowing to attack computationally more complex problems.

Symmetries are believed to be at heart of physics around us, including CPT: of charge conjugation (C) + parity transformation (P) + time reversal (T). \textbf{CPT theorem}, originally proven by Julian Schwinger~\cite{CPT}, is one of the reasons physicists generally believe this symmetry is satisfied by our physics, what was confirmed by many experimental tests~\cite{CPTdata}. 

In contrast, often considered \textbf{one-way quantum computers} (1WQC, also called measurement-based)~\cite{1WQC}, while using symmetric, unitary, reversible evolution through quantum gates, they treat the two boundary conditions in completely different ways. From one side there is state preparation (active), allowing to enforce initial states to e.g. $|0\rangle$ qubit values. However, from the second side there is available only measurement: taking a random value instead ($\textrm{Pr}(x_i)=\textrm{Tr}(\Pi_i \rho)$) - passively watching the outcome, instead of active influence.

Living in CPT symmetric physics suggests to consider enhancement by adding looking missing basic tool/operation: CPT analog of state preparation, denoted here as CPT(state preparation). For example using CPT analogs: pull/push, negative/positive pressure, stimulated emission/absorption causing deexcitation/excitation. Like in Fig. \ref{intr}, for hydrodynamics e.g. some future superfluid quantum computer~\cite{superfluid}, we could directly connect microfluid quantum chip into a circuit with pump: to actively both push into with positive pressure to enforce the initial states, and pull from with negative pressure to enforce the final states. 

As hydrodynamics and electrodynamics are mathematically similar, we will focus on looking most practical: photonic realizations, using stimulated emission/absorption for negative/positive radiation pressure. E.g. one of them could prepare/enforce $|1\rangle$ initial state. Therefore, the second should allow for its CPT analog: enforce $\langle 0|$ final state in more symmetric formulation of  quantum computation: through $\langle \Phi_{\textrm{final}}|U_{\textrm{quantum gates}}|\Phi_{\textrm{initial}}\rangle$ as in scattering theory~\cite{scat}.

This article is introduction to such proposed hypothetical possibility suggested in \cite{my}, and its potential realizations, applications to help achieving quantum supremacy by enhancing current approaches with CPT(state preparation) toward two-way quantum computers (2WQC). For example to restrict ensemble to satisfying some constraints $\psi=\sum_{a:\textrm{SAT}(a)} |a\rangle$, in theory allowing to attack general NP problems like 3-SAT in Fig. \ref{IQC}, \ref{2WQC}, maybe also helping with error correction.

\section{Two-way quantum computers (2WQC)}
\begin{figure}[t!]
    \centering
        \includegraphics[width=9cm]{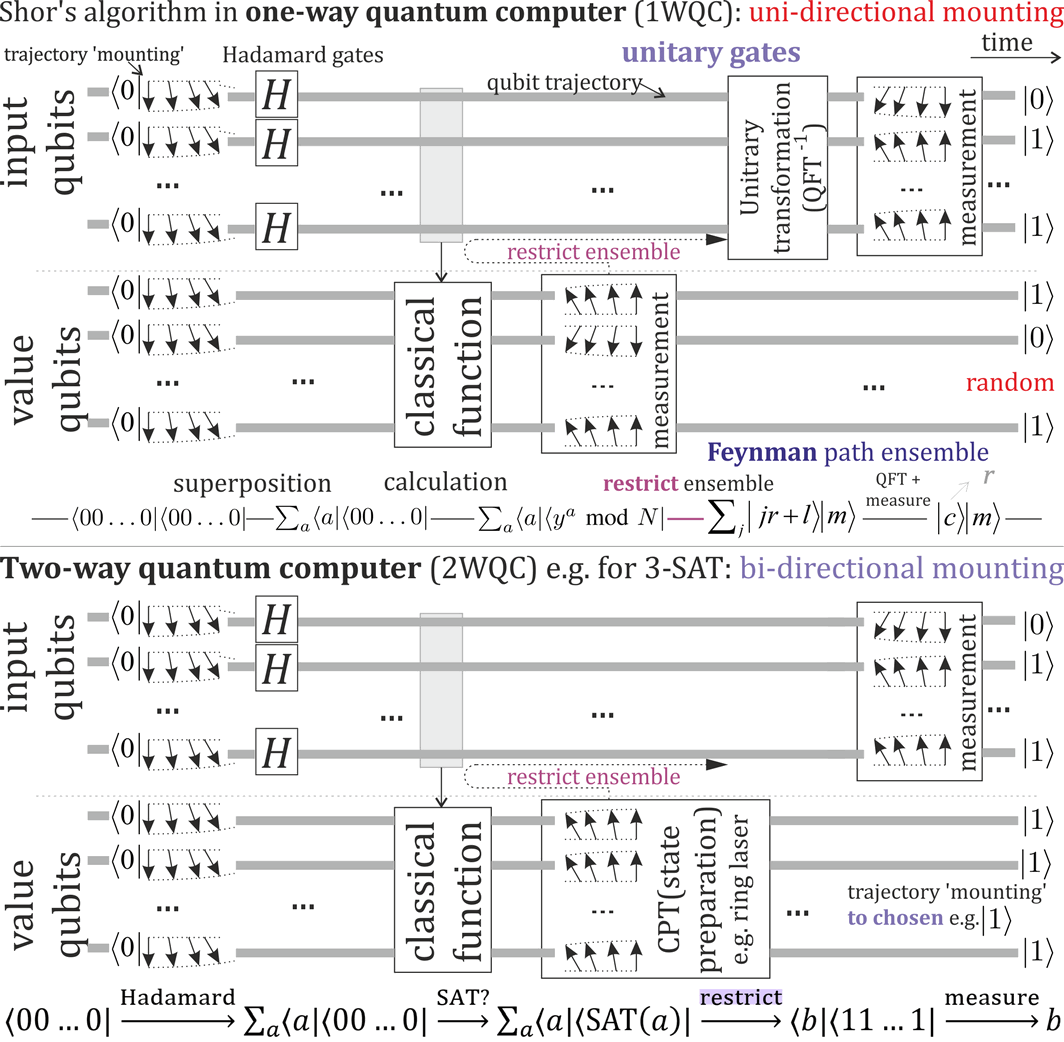}
        \caption{ \textbf{Top}: Schematic diagram of quantum subroutine of Shor's algorithm~\cite{shor} for finding prime factors of natural number $N$, for convenience using  $\langle \Phi_{\textrm{initial}}|U_{\textrm{quantum gates}}|\Phi_{\textrm{final}}\rangle$ convention. For a random natural number $y<N$, it searches for period $r$ of $f(a)=y^a \textrm{ mod }N$ classical function. This period can be concluded from measurement of value $c$ after Quantum Fourier Transform ($\textrm{QFT}^{-1}$) and with some large probability $(O(1))$ allows to find a nontrivial factor of $N$. The Hadamard gates produce state being superposition of all possible values of $a$: $2^n$ for length $n$. Then classical function $f(a)$ is applied, getting entanglement  $\sum_a |a\rangle |f(a)\rangle$. Due to necessary reversibility of applied operations, this calculation of $f(a)$. Now measuring the value of $f(a)$ returns some random value $m$, and restricts the original superposition to only $a$ fulfilling $f(a)=m$. Mathematics ensures that $\{a:f(a)=m\}$ set has to be periodic here $(y^r \equiv 1 \mod N)$, this period $r$ is concluded from the value of Fourier Transform ($\textrm{QFT}^{-1}$). 
        \textbf{Bottom}: Analogous example approach to two-way quantum computer (2WQC): adding CPT(state preparation) to enforce some values of classical function, e.g. of verifier testing satisfaction of some constraints like alternatives in 3-SAT problems. }       \label{IQC}
\end{figure}

Standard one-way quantum computers (1WQC), also called measurement-based, can start with state preparation, then use Hadamard gates to prepare entanglement of $2^n$ inputs, then e.g. calculation of some classical function, Quantum Fourier Transform (QFT), and finally measurements. Like in Fig. \ref{IQC} - for Shor algorithm~\cite{shor} measurement of calculated classical function restricts the ensemble to often periodic subset. Then QFT and measurement allows to obtain this period, providing hint for the factorization problem.

This factorization method was introduced by Peter Shor in 1994, and is still one of the most promising quantum algorithm. However, there is a general belief that current quantum computers cannot attack general NP problems - what in theory could be reachable by their proposed enhancement from 1WQC to 2WQC by more precise (bidirectional) control of information flow.

\subsection{NP problems}
We will focus on NP (nondeterministic polynomial) problems~\cite{np}: in which there is a verifier allowing to test in a polynomial time if a given input is correct, and we would like to find such satisfying input(s), or just test if it exists. The difficulty is in exponential number of inputs, e.g. $2^n$ for $n$ bit input. Mentioned factorization problem can be seen this way: verifier would test if a given input divides some fixed number.  

In contrast, P problems are those solvable in a polynomial time on a classical computer. While P $\subset$ NP, there is fundamental open  "P vs NP" question of their equality. Using standard assumption that such polynomial time algorithms for all NP problems do not exist, there are defined \textbf{NP-complete problems} as the most difficult subfamily with polynomial transformation between each other, e.g. 3-SAT problem. This way finding polynomial time algorithm for one of them, or proving it does not exist, we would do it for all NP-complete problems.

While there are known lots of NP-complete problems, for simplicity (also to implement) let us focus on \textbf{3-SAT}. For multiple alternatives of 3 binary variables (some can be negated), the question is if we can choose values for all variables to satisfy all such alternatives, e.g.:
$$\exists_{x_1 x_2 \ldots}\ (x_1 \vee\neg x_2 \vee x_3) \wedge (\neg x_4 \vee x_2 \vee \neg x_3) \wedge (x_5 \vee \neg x_4 \vee x_2)\wedge \ldots\ ?$$
Allowing to additionally also use "for all" $\forall$ universal quantifier, we would get to PSPACE-complete problems~\cite{pspace}: analogous family for solvable in polynomial space (PSPACE), generally believed to be even more difficult - hence maybe worth to consider for post-quantum cryptography.
 
For standard quantum computers there is usually considered BQP (bounded-error quantum polynomial time complexity)~\cite{bqp}: which can be solved e.g. by 1WQC in polynomial time, with an error probability of at most 1/3 for all instances. Thanks to 1994 Shor algorithm as in Fig. \ref{IQC}, factorization problem is example of problems being in BQP, but believed not to be in P - giving hope for superiority of quantum computers. 

It is also believed that NP-complete problems are outside BQP. However, in Ising model we can easily formulate various NP-complete problems~\cite{ising} - e.g. such that perfect Boltzmann ensemble would lead to a configuration satisfying all the constraints (verifier). While quantum computers use mathematically similar Feynman path ensemble instead, their weakness here is in the measurement: taking a random value, instead of fixing a chosen one like for state preparation (in Ising we can symmetrically fix both sides) - they are missing CPT(state preparation), we would like to add here.

\subsection{2WQC approach for NP-complete problems }
While state preparation allows to enforce physical constraints on the initial state, having also symmetric CPT(state preparation) to simultaneously enforce physical constraints on some final states, we could e.g. try to enforce satisfaction of verifier, constraints of a given NP problem, hopefully restricting the ensemble to those satisfying these constraints. 

For example as in Fig. \ref{IQC}: like in Shor algorithm splitting variables as $|\textrm{input} \rangle_1 |\textrm{value} \rangle_2$ both prepared as $|00\ldots 0 \rangle$. Apply Hadamard gates to input bits to get ensemble of all inputs. Then calculate constraints for input into value variables, e.g. alternatives for 3-SAT: using NOT and Controlled-OR gates. 

Then we would like to enforce outcomes of these alternatives to '1' (true) by applying CPT(state preparation) to all as $\langle 11\ldots 1|$ e.g. using negative pressure, ring laser. This way we should prepare:
$$\psi=\langle 11\ldots 1|_2 (\textrm{Controlled-(N)ORs})(H \otimes I) |00\ldots 0 \rangle_1 |00\ldots 0 \rangle_2 $$ 
with restricted ensemble ideally to satisfying the 3-SAT instance: $\psi=\sum_{a:\textrm{SAT}(a)} |a\rangle$. Measuring its qubits, we should obtain one of these satisfying inputs. If there is no satisfying input, imperfections should lead e.g. to some random values.

Defining 2WBQP in analogy to BQP for 2WQC - additionally allowing for CPT(state preparation) operation, we could calculate verifier (into values) and enforce its satisfaction like for 3-SAT, suggesting NP $\subset$ 2WBQP. From the other side 2WBQP $\subset$ PSPACE. It is interesting open complexity question if it could solve a larger class than NP, e.g. by performing some additional operations on such $\psi=\sum_{a:\textrm{verified}(a)} |a\rangle$ ensemble.  

However, there will be rather unavoidable \textbf{imperfections}, for example modelled like in binary symmetric channel: instead of perfect $\langle 0|$ there would be $\sqrt{1-\epsilon^2}\,\langle 0|+\epsilon \langle 1|$ final state, and $\epsilon \langle 0|+\sqrt{1-\epsilon^2}\, \langle 1|$ instead of $\langle 1|$, for some small $\epsilon >0$. We could group multiple such imperfect prepared qubits and use some error correction technique e.g. majority voting as the final prepared qubit, reducing $\epsilon$ to arbitrarily low. 

There are also other imperfections, which might prevent practical attacks on NP-complete problems, what requires deeper analysis. However, there are already many claims of quantum superiority, close to which enhancements with such additional new tool as even imperfect CPT(state preparation) should help with improvements, maybe also qualitatively: toward additional computational classes like NP-complete. From the other side, this additional operation could help stabilizing behavior - e.g. to help with error correction also of 1WQC.

\begin{figure}[t!]
    \centering
        \includegraphics[width=9cm]{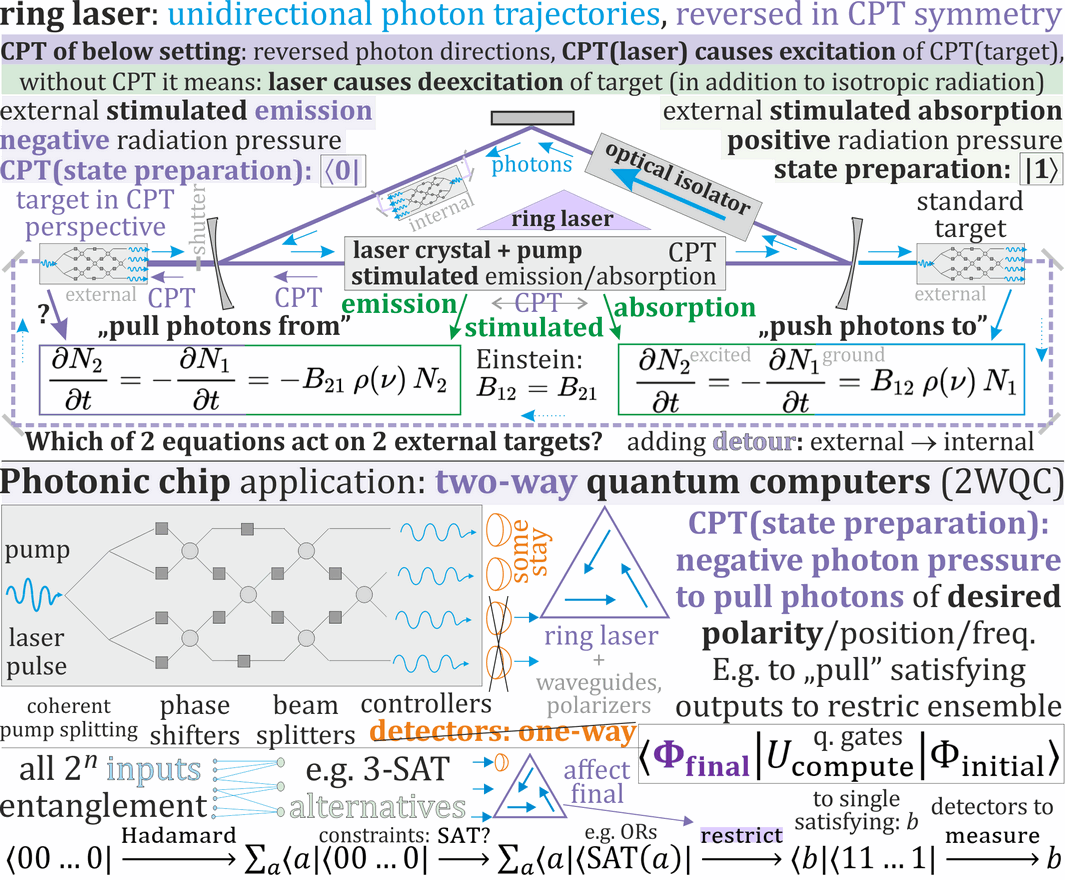}
        \caption{\textbf{Top}: stimulated emission/absorption can be seen as CPT analogs, governed by the written equations for $N_1$/$N_2$ numbers of atoms in the ground/excited state. One can be used for state preparation, suggesting to try to use the second for CPT(state preparation) - e.g. ring laser (unidirectional photon trajectories) as pump for photons, and internal placement of \textbf{photonic chip: such that both equations act on it} (e.g. with switchable mirrors for pulsed action or shown detour) to both push and pull photons through chip for better control. 
        \textbf{Bottom}: example of application for photonic 2WQC - seeing quantum computation as $\langle \Phi_{\textrm{final}}|U_{\textrm{quantum gates}}|\Phi_{\textrm{initial}}\rangle$ with affected both states: initial by state preparation and final by CPT(state preparation). Shown approach prepares ensemble of all inputs, calculates constraints e.g. alternatives for SAT, then use ring laser for CPT(state preparation) to enforce '1' for satisfaction of all these constraints - hopefully restricting ensemble to  $\sum_{a:\textrm{SAT}(a)} |a\rangle$, measurement should extract one of them.
        }       \label{2WQC}
\end{figure}

\section{Potential physical realizations} 
For hydrodynamics and state preparation as pushing with positive pressure like in Fig. \ref{intr}, its CPT analog would be trivial: just pulling with negative pressure from the opposite side. We could do both simultaneously e.g. connecting microfluidic chip into a circuit with pump. Building quantum computer this way seems very difficult, but might be possible in some future with superfluids~\cite{superfluid}. Fortunately, hydrodynamics and electromagnetism are governed by similar wavelike PDEs, hopefully allowing also for other approaches, e.g. EM microwave~\cite{micro} or preferably photonic - we will focus on.

It is natural to heat up or push with photons - carriers of energy,  momentum and angular momentum. Existence of CPT analogs of these tasks probably seemed impossible in the past, however, they turned out possible and realizable - in currently popular \textbf{laser cooling}~\cite{cooling}, and \textbf{optical tweezers}~\cite{tweezers} - both awarded with Nobel prizes (1997 for Claude Cohen-Tannoudji, Steven Chu, William Daniel Phillips and 2018 for Arthur Ashkin). The latter allows for \textbf{optical pulling}, realized with various approaches, e.g. \cite{pull1,pull}.

EM radiation pressure is a vector $\vec{p}=\langle \vec{E}\times \vec{H} \rangle/c$: which is not necessarily toward given surface, allowing for both positive (toward) and \textbf{negative radiation pressure} (outward) - the latter is considered to pull e.g. solitons~(\cite{neg1,neg2}). Pushing/pulling with positive/negative radiation pressure can be seen as CPT analogs, and for photon flux has seeming related \textbf{stimulated emission-absorption} equations being at heart of lasers, governing deexcitation-excitation ($N_1$ ground state, $N_2$ excited atoms):

\be\textrm{stimulated emission:}\ \frac{\partial N_2}{\partial t}=-\frac{\partial N_1}{\partial t}=-B_{21}\,\rho(\nu)\, N_2\label{ee}\ee
\be\textrm{and absorption:}\ \frac{\partial N_2}{\partial t}=-\frac{\partial N_1}{\partial t}=B_{12}\,\rho(\nu)\, N_1 \label{ae}\ee 
\noindent where $B_{12}=B_{21}$ are (symmetric) Einstein's coefficients~\cite{Einstein}, $\rho(\nu)$ is radiation density of the incident field at frequency $\nu$, corresponding to transition between the two considered states: $E_2-E_1=h\nu$.

For \textbf{laser}~\cite{lasers} interior it is assumed both these equations act on active laser medium e.g. pumped crystal. Adding them we get domination of deexcitation for $\Delta N = N_2-N_1 > 0 $ population inversion achieved e.g. in lasers. Additionally, there is spontaneous emission $\partial N_2/\partial t=-\partial N_1/\partial t=-A_{21}\,N_2$, which is time asymmetric - what can be imagined through properties of solution we live in: while it is easy for photons to escape e.g. to be absorbed in the future, its T symmetric analog would require (uncommon) their source in the past.

While in standard laser photons travel in both directions, there are also available \textbf{ring lasers} with circular photon trajectories, and often (assumed here) optical isolator enforcing nearly unidirectional photon trajectories. It is usually achieved by the Faraday effect: difference of propagation speed for two circular polarizations, leading to rotation of direction of linear polarization. Adding two linear polarizers, we can allow for only photons in one direction. Applying T symmetry would reverse photon trajectories, because it would exchange the two circular polarizations, hence such materials violate T symmetry~\cite{faraday}.

\subsection{Stimulated emission-absorption for external targets}
While for internal objects like the active laser medium it is assumed that both (\ref{ee})+(\ref{ae}) equations act on them, for external targets a  search (literature, discussion forums, private communication) suggests there remains open fundamental question regarding the stimulated emission equation (\ref{ee}).

The difference between internal and external targets is in photon flux: coming from both (internal) or one (external) direction. Looking from perspective with applied CPT symmetry, photon trajectory would be reversed, e.g. for ring laser as in Fig. \ref{2WQC} leading to additional usually ignored targets - on which there should act (\ref{ae}) absorption equation, what in standard perspective (no CPT) means acting (\ref{ee}) emission equation. Without preparation $N_2 \approx 0$, making such effect negligible. 

Adding detour as in this diagram, would transform both external targets into internal - both equations should act on them. However, the detour directly adds only absorption equation (\ref{ae}) to target on the left, again suggesting the emission equation (\ref{ee}) was already acting on it without detour.

Therefore, let us formulate two alternative hypotheses: HCPT assuming CPT symmetry, and HAO violating it:

\textbf{Hypothesis CPT [HCPT]}: stimulated emission equation (\ref{ee}) acts on targets of reversed photon trajectories (because absorption equation (\ref{ae}) acts on it in CPT perspective).

\textbf{Hypothesis absorption only [HAO]}: only (\ref{ae}) absorption equation applies to external targets.

It seems an open question which one is true, there might be also some different e.g. intermediate possibilities. For example belief in impossibility of reaching population inversion for two-state systems, or "Asking photons where they have been"~\cite{asking} article experiment requiring photon paths in both time direction, suggest HCPT. Here are some examples of possible direct \textbf{experimental tests HCPT vs HAO}:
\begin{itemize}
  \item Prepare conditions that only stimulated emission (\ref{ee}) equation should act on a given target, like behind ring laser in Fig. \ref{2WQC}, and maintain excitation $N_2>0$ of this target e.g. through some external pumping. Monitoring its population level, and opening a shutter toward such laser - if it increases deexcitation rate accordingly to emission equation (\ref{ae}), then HCPT is true, otherwise HAO.
  \item In standard laser (no optical isolator) photons travel in both directions, suggesting that both (\ref{ee})+(\ref{ae}) equations should act on external targets, what would bound the maximal excited population of such two-state external target to $N_2 \leq N_1$ for HCPT hypothesis. In contrast, for HAO it might be possible to achieve population inversion $N_2>N_1$ (if overcoming spontaneous emission). For HCPT shooting ring laser could allow to reach population inversion for two-state system.
  \item Analogously if for standard laser both equations act on external target (HCPT), would mean adding tendency toward $N_2=N_1$: not only increasing $N_2$, but decreasing instead in case of $N_2>N_1$ population inversion - such decrease would confirm HCPT, its lack HAO.
\end{itemize}
More direct test for 2WQC would be placing beamsplitter e.g. in detour of ring laser as in top-right of Fig. \ref{intr}, and testing if negative radiation pressure reduces flow from beamsplitter (down) e.g. by opening/closing shutter in this direction.

\subsection{Photonic quantum computer application}

Photonic quantum computer approaches often (e.g. \cite{ph1,ph2,ph3,ph4}) use external pulsed light for pumping (state preparation), followed by SFWM (spontaneous four-wave mixing) to generate entangled photons. Such pumping can be seen as using the behavior from  absorption equation (\ref{ae}). From the other direction of such photonic chip there are used detectors - in 2WQC some of which we would like to replace with CPT(state preparation): using behavior from the stimulated emission equation (\ref{ee}), e.g. in pulsed way - later than initial pulse by time of propagation through the chip.

In case of HCPT we should be able to achieve it for external target behind ring laser as in Fig. \ref{2WQC}. However, in both cases (also HAO), there should remain possibility by internal placement of photonic chip - inside the unidirectional photon flow of ring laser (or adding detour), where it is generally believed that both (\ref{ee})+(\ref{ae}) equations apply. A system of e.g. switchable mirrors should allow to redirect ring laser photon flux as pulses through photonic chip - for better controlled by both pushing and pulling radiation through chip.

\section{Conclusions and further work}
This is introductory theoretical article proposing basic ideas for 2WQC - hopefully leading to experimental verifications, realizations, enhancements of current quantum computers with CPT(state preparation) to help achieving quantum supremacy.

Here are some theoretical questions to be explored e.g. in future versions of this article:
\begin{itemize}
  \item For basic CPT analogs of state preparation there was discussed application of very popular stimulated emission-absorption equations, with focus on straightforward application for photonic quantum computers. However, there are also many other approaches for quantum computers worth to confider for 2WQC in the future - e.g. trapped ions~\cite{qc1}, neutral atoms~\cite{qc2}, superconducting loops~\cite{qc3}, silicon quantum dots~\cite{qc4}, topological qubits~\cite{qc5}, and diamond vacancies~\cite{qc6}. For some of them there could be also used stimulated emission-absorption analogs, e.g. as pumping to $|1\rangle$ - "unpumping" to $\langle 0|$. For others e.g. some forms of positive/negative radiation pressure might be considered, like to both push and pull topological defects through chip realized in superconductor. For topological qubits measurement is e.g. through annihilation - choosing its target might allow to enforce the final state. 
  \item For practical realizations, imperfections in all levels need to be included into considerations, here with addition from likely very imperfect CPT(state preparation), e.g. treated as having tiny contributions of opposite values. It requires additional e.g. error correction mechanisms, which like for Shor algorithm might essentially restrict practical possibilities. It requires investigation, also search for other ways CPT(state preparation) could enhance e.g. current approaches to help achieving quantum supremacy, maybe also used to just help with error correction by stabilizing the flow through bidirectional control.
  \item While there was discussed 3-SAT as example of NP-complete problems, similar approach using ensemble restriction by CPT(state preparation) enforcing constraints should be applicable to other NP-complete problems. There might be also different 2WQC approaches to search for and investigate, e.g. reversing unitary computation. There is polynomial equivalence among NP-complete problems, suggesting search for the optimal ones for various specific tasks, also taking into consideration quantum computer architecture, imperfections/error correction, etc.
  \item While in theory 2WQC could allow to attack NP problems like 3-SAT, there remains open question of solving PSPACE problems in polynomial time, like quantified Boolean formula problem~\cite{pspace} adding universal quantifier $\forall$ to SAT formulas. There might be some intermediate class between NP and PSPACE of achievable problems, its characterization is interesting open problem.
  \item There are already claims of quantum superiority, close to them addition of CPT(state preparation) might allow for essential improvements - worth search and investigation.        
  \item Such CPT(state preparation) might also lead to other possibilities to explore. For example having controlled-XOR gate: $(x,y,z) \to (x,y, (x\  \textrm{XOR}\ y)\ \textrm{XOR}\ z)$, and fixing its last variable to zero in both directions: $\langle 0| z|0\rangle$ e.g. through placing it inside ring laser, should enforce $x=y$. Using it multiple times on past-future zigzag, should enforce equality of far away variables, potentially allowing for faster than light communication - hydrodynamics is too slow for that (speed of sound), but it seems worth testing for EM/photonics (speed of light).
  \item As in some future there might appear attacks on NP problems, it seems worth to search for PSPACE-based cryptography, e.g. through some game between authorising devices, or reconfiguration problems requiring to find a path e.g. of given key-dependant hash values.
\end{itemize}

\bibliographystyle{IEEEtran}
\bibliography{cites}

\begin{thebibliography}{10}
\providecommand{\url}[1]{#1}
\csname url@samestyle\endcsname
\providecommand{\newblock}{\relax}
\providecommand{\bibinfo}[2]{#2}
\providecommand{\BIBentrySTDinterwordspacing}{\spaceskip=0pt\relax}
\providecommand{\BIBentryALTinterwordstretchfactor}{4}
\providecommand{\BIBentryALTinterwordspacing}{\spaceskip=\fontdimen2\font plus
\BIBentryALTinterwordstretchfactor\fontdimen3\font minus
  \fontdimen4\font\relax}
\providecommand{\BIBforeignlanguage}[2]{{%
\expandafter\ifx\csname l@#1\endcsname\relax
\typeout{** WARNING: IEEEtran.bst: No hyphenation pattern has been}%
\typeout{** loaded for the language `#1'. Using the pattern for}%
\typeout{** the default language instead.}%
\else
\language=\csname l@#1\endcsname
\fi
#2}}
\providecommand{\BIBdecl}{\relax}
\BIBdecl

\bibitem{EMh}
A.~I. Arbab, ``The analogy between electromagnetism and hydrodynamics,''
  \emph{Physics Essays}, vol.~24, no.~2, p. 254, 2011.

\bibitem{fluid}
P.~Cui and S.~Wang, ``Application of microfluidic chip technology in
  pharmaceutical analysis: A review,'' \emph{Journal of pharmaceutical
  analysis}, vol.~9, no.~4, pp. 238--247, 2019.

\bibitem{CPT}
J.~Schwinger, ``The theory of quantized fields. i,'' \emph{Physical Review},
  vol.~82, no.~6, p. 914, 1951.

\bibitem{CPTdata}
V.~A. Kosteleck{\`y} and N.~Russell, ``Data tables for lorentz and c p t
  violation,'' \emph{Reviews of Modern Physics}, vol.~83, no.~1, p.~11, 2011.

\bibitem{1WQC}
R.~Raussendorf and H.~J. Briegel, ``A one-way quantum computer,''
  \emph{Physical review letters}, vol.~86, no.~22, p. 5188, 2001.

\bibitem{qc1}
K.~R. Brown, J.~Chiaverini, J.~M. Sage, and H.~H{\"a}ffner, ``Materials
  challenges for trapped-ion quantum computers,'' \emph{Nature Reviews
  Materials}, vol.~6, no.~10, pp. 892--905, 2021.

\bibitem{qc2}
D.~S. Weiss and M.~Saffman, ``Quantum computing with neutral atoms,''
  \emph{Physics Today}, vol.~70, no.~7, pp. 44--50, 2017.

\bibitem{qc3}
S.~Kwon, A.~Tomonaga, G.~Lakshmi~Bhai, S.~J. Devitt, and J.-S. Tsai,
  ``Gate-based superconducting quantum computing,'' \emph{Journal of Applied
  Physics}, vol. 129, no.~4, p. 041102, 2021.

\bibitem{qc4}
M.~Vinet, L.~Hutin, B.~Bertrand, S.~Barraud, J.-M. Hartmann, Y.-J. Kim,
  V.~Mazzocchi, A.~Amisse, H.~Bohuslavskyi, L.~Bourdet \emph{et~al.}, ``Towards
  scalable silicon quantum computing,'' in \emph{2018 IEEE International
  Electron Devices Meeting (IEDM)}.\hskip 1em plus 0.5em minus 0.4em\relax
  IEEE, 2018, pp. 6--5.

\bibitem{superfluid}
Y.~L. Sfendla, C.~G. Baker, G.~I. Harris, L.~Tian, R.~A. Harrison, and W.~P.
  Bowen, ``Extreme quantum nonlinearity in superfluid thin-film surface
  waves,'' \emph{npj Quantum Information}, vol.~7, no.~1, p.~62, 2021.

\bibitem{phonon}
H.~Qiao, {\'E}.~Dumur, G.~Andersson, H.~Yan, M.-H. Chou, J.~Grebel, C.~Conner,
  Y.~Joshi, J.~Miller, R.~Povey \emph{et~al.}, ``Splitting phonons: Building a
  platform for linear mechanical quantum computing,'' \emph{Science}, vol. 380,
  no. 6649, pp. 1030--1033, 2023.

\bibitem{micro}
J.~C. Bardin, D.~H. Slichter, and D.~J. Reilly, ``Microwaves in quantum
  computing,'' \emph{IEEE journal of microwaves}, vol.~1, no.~1, pp. 403--427,
  2021.

\bibitem{scat}
W.~Greiner and J.~Reinhardt, \emph{Field quantization}.\hskip 1em plus 0.5em
  minus 0.4em\relax Springer Science \& Business Media, 1996.

\bibitem{my}
J.~Duda, ``Four-dimensional understanding of quantum mechanics,'' \emph{arXiv
  preprint arXiv:0910.2724}, 2009.

\bibitem{shor}
P.~W. Shor, ``Polynomial-time algorithms for prime factorization and discrete
  logarithms on a quantum computer,'' \emph{SIAM review}, vol.~41, no.~2, pp.
  303--332, 1999.

\bibitem{np}
L.~Fortnow, ``The status of the p versus np problem,'' \emph{Communications of
  the ACM}, vol.~52, no.~9, pp. 78--86, 2009.

\bibitem{pspace}
A.~Fl{\"o}gel, M.~Karpinski, and H.~K. B{\"u}ning, ``Subclasses of quantified
  boolean formulas,'' in \emph{International Workshop on Computer Science
  Logic}.\hskip 1em plus 0.5em minus 0.4em\relax Springer, 1990, pp. 145--155.

\bibitem{bqp}
S.~Aaronson, ``Bqp and the polynomial hierarchy,'' in \emph{Proceedings of the
  forty-second ACM symposium on Theory of computing}, 2010, pp. 141--150.

\bibitem{ising}
A.~Lucas, ``Ising formulations of many np problems,'' \emph{Frontiers in
  physics}, vol.~2, p.~5, 2014.

\bibitem{cooling}
C.~N. Cohen-Tannoudji and W.~D. Phillips, ``New mechanisms for laser cooling,''
  \emph{Physics Today}, vol.~43, no.~10, pp. 33--40, 1990.

\bibitem{tweezers}
A.~Ashkin, ``Acceleration and trapping of particles by radiation pressure,''
  \emph{Physical review letters}, vol.~24, no.~4, p. 156, 1970.

\bibitem{pull1}
J.~Chen, J.~Ng, Z.~Lin, and C.~T. Chan, ``Optical pulling force,'' \emph{Nature
  photonics}, vol.~5, no.~9, pp. 531--534, 2011.

\bibitem{pull}
L.~Wang, S.~Wang, Q.~Zhao, and X.~Wang, ``Macroscopic laser pulling based on
  the knudsen force in rarefied gas,'' \emph{Optics Express}, vol.~31, no.~2,
  pp. 2665--2674, 2023.

\bibitem{neg1}
P.~Forg{\'a}cs, {\'A}.~Luk{\'a}cs, and T.~Roma{\'n}czukiewicz, ``Negative
  radiation pressure exerted on kinks,'' \emph{Physical Review D}, vol.~77,
  no.~12, p. 125012, 2008.

\bibitem{neg2}
A.~Mizrahi and Y.~Fainman, ``Negative radiation pressure on gain medium
  structures,'' \emph{Optics letters}, vol.~35, no.~20, pp. 3405--3407, 2010.

\bibitem{Einstein}
A.~Einstein, ``Strahlungs-emission und-absorption nach der quantentheorie, 17
  jul 1916,'' 1916.

\bibitem{lasers}
A.~E. Siegman, \emph{Lasers}.\hskip 1em plus 0.5em minus 0.4em\relax University
  science books, 1986.

\bibitem{faraday}
O.~Sigwarth and C.~Miniatura, ``Time reversal and reciprocity,'' \emph{AAPPS
  Bulletin}, vol.~32, no.~1, p.~23, 2022.

\bibitem{asking}
A.~Danan, D.~Farfurnik, S.~Bar-Ad, and L.~Vaidman, ``Asking photons where they
  have been,'' \emph{Physical review letters}, vol. 111, no.~24, p. 240402,
  2013.

\bibitem{ph1}
J.~Wang, D.~Bonneau, M.~Villa, J.~W. Silverstone, R.~Santagati, S.~Miki,
  T.~Yamashita, M.~Fujiwara, M.~Sasaki, H.~Terai \emph{et~al.}, ``Chip-to-chip
  quantum photonic interconnect by path-polarization interconversion,''
  \emph{Optica}, vol.~3, no.~4, pp. 407--413, 2016.

\bibitem{ph2}
J.~Wang, S.~Paesani, Y.~Ding, R.~Santagati, P.~Skrzypczyk, A.~Salavrakos,
  J.~Tura, R.~Augusiak, L.~Man{\v{c}}inska, D.~Bacco \emph{et~al.},
  ``Multidimensional quantum entanglement with large-scale integrated optics,''
  \emph{Science}, vol. 360, no. 6386, pp. 285--291, 2018.

\bibitem{ph3}
J.~C. Adcock, C.~Vigliar, R.~Santagati, J.~W. Silverstone, and M.~G. Thompson,
  ``Programmable four-photon graph states on a silicon chip,'' \emph{Nature
  communications}, vol.~10, no.~1, p. 3528, 2019.

\bibitem{ph4}
D.~Llewellyn, Y.~Ding, I.~I. Faruque, S.~Paesani, D.~Bacco, R.~Santagati, Y.-J.
  Qian, Y.~Li, Y.-F. Xiao, M.~Huber \emph{et~al.}, ``Chip-to-chip quantum
  teleportation and multi-photon entanglement in silicon,'' \emph{Nature
  Physics}, vol.~16, no.~2, pp. 148--153, 2020.

\bibitem{qc5}
A.~Stern and N.~H. Lindner, ``Topological quantum computation—from basic
  concepts to first experiments,'' \emph{Science}, vol. 339, no. 6124, pp.
  1179--1184, 2013.

\bibitem{qc6}
S.~Pezzagna and J.~Meijer, ``Quantum computer based on color centers in
  diamond,'' \emph{Applied Physics Reviews}, vol.~8, no.~1, 2021.

\end{thebibliography}
\end{document}